\documentclass[11pt]{article}
\usepackage{graphics}

\newfont{\headfont}{cmbx10 scaled 1440}
\newfont{\headfontb}{cmbx10 scaled 1200}
\newfont{\namefont}{cmr9}
\newfont{\initialfont}{cmr9 scaled 1200}
\newfont{\addfont}{cmti10}

\begin{document}
\baselineskip=.41cm
\begin{center}
{\headfont Decoherence of the Unruh Detector}
\end{center}
\vskip 0.05truein
\begin{center}
{ {\initialfont J}{\namefont EAN-}{\initialfont G}{\namefont UY}
                    {\initialfont D}{\namefont EMERS   }}
\end{center}
\begin{center}
{\addfont{Department of Physics,McGill University}}\\
{\addfont{3600 University St., Montreal, PQ, Canada H3A 2T8.} }\\
{\addfont{ jgdemers@hep.physics.mcgill.ca} }\\

\vskip 0.1truein
 {\bf Abstract}
\end{center}
\vspace{-0.8cm}
\begin{quote}
As it is well known, the Minkowski vacuum appears thermally populated to a
quantum mechanical detector on a uniformly accelerating course. We investigate
 how this thermal radiation may contribute to the classical nature of the
detector's
trajectory through the criteria of decoherence. An uncertainty-type relation is
obtained
 for the detector   involving the fluctuation in temperature, the time of
flight
and the coupling to the bath.
\end{quote}

\vspace{-1.0cm}
\section{Introduction}\label{intro}
In the Copenhagen interpretation of the measurement process,
the existence of classical systems is postulated from the start.
In quantum cosmology however, the whole universe receives a quantum mechanical
description
and classicality must be somehow derived. An important mechanism by which some
degrees
of freedom may behave classically is the environment-induced decoherence. In
that case,
partial diagonalization of the density matrix in a given  basis is obtained by
integrating
out some unobserved  degrees of freedom\cite{Zu}.
A simple illustration is given by
a system and its environment having  each a single degree
of freedom  described by the actions
$S_{sys}[x]$ and $S_{env}[y]$ respectively,  with interaction
  described by the action $S_{int}[x,y]$.
Starting with  an uncorrelated   density matrix  at   $t=0$,
\begin{equation}
\rho(x_1,y_1,x_2,y_2;t=0)=\rho_{sys}(x_1,x_2;t=0)\times
\rho_{env}(y_1,y_2;t=0),
\end{equation}
then at a time $t$ later, the reduced density matrix (where the
environment is integrated out) will be  given by\cite{Hal,Hu1}:
\begin{equation}
\rho_{red}(x_1^f,x_2^f;t)=\int dx_1^i\:\int dx_2^i\:
P(x_1^f,x_2^f,t|x_1^i,x_2^i,0)\:\rho_{sys}(x_1^i,x_2^i;0),
\end{equation}
with the kernel
\begin{equation}
P(x_1^f,x_2^f,t|x_1^i,x_2^i,0)=\int_{x_1^i}^{x_1^f}
{\cal D} x_1\:\int_{x_2^i}^{x_2^f} {\cal D} x_2  \:e^{i(S_{sys}[x_1]-
S_{sys}[x_2])}\:\:e^{i\Gamma[x_1,x_2]},
\label{p}
\end{equation}
where   the  boundary conditions on the functional integrals
are given for  times $0$ and $t$ (our units are such
that $\hbar=c=1$).
$\Gamma[x_1,x_2]$ is the influence functional (IF).
It is a property of the environment and  the way it is coupled to the
system\cite{Fey1}.
For an environment  in a pure state at $t=0$,  it
takes the conceptually simple form
\begin{equation}
e^{i\Gamma[x_1,x_2]}=\langle\psi_2(t)|\psi_1(t)\rangle.
\label{if2}
\end{equation}
Here, $|\psi_1(t)\rangle$ is the state that   evolved from the
initial state at $t=0$ under the dynamics dictated  by
$S_{env}[y_1]+S_{int}[x_1,y_1]$ in which  $x_1$ acts  as a c-number,
time dependent source; likewise   $|\psi_2(t)\rangle$ is
governed by $S_{env}[y_2]+S_{int}[x_2,y_2]$.
For    paths $[x_1(t)]$ and  $[x_2(t)]$   such that the
states in (\ref{if2}) are not adiabatically disturbed, $\Gamma$ will
have   a (positive)
imaginary part   $\Gamma_I$.
It is clear that
if  $\Gamma_I$ is large for   pair of paths $(x_1(t), x_2(t))$ that
are far apart from one another in spacetime, then the contribution
of these pairs   will be suppressed in (\ref{p}). As a result,
$\rho_{red}(x_1^f,x_2^f;t)$ will  be more diagonal.  Decoherence
can thus be studied   through $\Gamma_I$.

In recent times, the study  of black holes received renewed interest, in part
du to
the introduction of more tractable two dimensional models.
An important issue arising in that context is the limit of   the semiclassical
approximation,
in which the Hawking radiation is derived.
As part of the groundwork to that problem, we report here on the related issue
of
the decoherence of a detector in uniform acceleration.
In the next section, we review how the IF predicts the heating
of the  scalar field forming the  environment of a detector
on a uniformly accelerating course\cite{Un,Anglin}.
In Section~\ref{fluc},
the spacetime   trajectory is  taken to be a {\it decohered}
spacetime path, with a spread around the uniformly accelerating
trajectory,
and taking now the detector's excitation to be fixed.
We  then obtain an uncertainty-type relation  involving  the spread in
acceleration  involved in
constructing the approximately classical path, the time of flight and the
coupling to the thermal bath.

\section{Unruh Effect}
\label{Unruh}
Consider an  environment made of a massless scalar field in
two spacetime dimensions\cite{Hu1,Anglin}. In a finite box
of   spatial dimension $L$, the mode expansion reads
\begin{equation}
\phi(x,t)=\sqrt{2\over L} \sum_k[y_k^+cos\: kx+y_k^-sin\: kx],
\label{mode}
\end{equation}
with $k=2\pi n/L$ with $n=1,2...$.
The kinetic term is
\begin{equation}
L_{env}=\int dx \: {1\over 2} \partial_{\mu}
\phi\:\partial ^\mu \phi ={1\over 2}\sum_{\sigma
=+,-}\sum_k[(\dot{y}_k^\sigma)^2 -k^2
( y _k^\sigma)^2  ] ,
\label{lenv}
\end{equation}
where dots denote time derivative. The system is formed by a
detector that is linearly and locally sensitive to the matter
field with an interacting Lagrangian density
\begin{equation}
{\cal L}_{int} (x) = -\varepsilon Q\phi(x,t) \delta(x-X(t)), \label{inte}
\end{equation}
 where $\varepsilon$ is a coupling constant while  $Q(t)$ is the
DeWitt monopole moment\cite{DeW} and  plays the role of $x(t)$
in Section~\ref{intro} while $X(t)$ is the trajectory of the detector.
For a uniform acceleration $a$,
$X(\tau)=\frac{1}{a}\cosh(a\tau)$ and $t(\tau)=\frac{1}{a}\sinh(a\tau)$ where
$\tau$ is the proper time and the resulting IF is given by \cite{Hu1}
\begin{eqnarray}
\Gamma_R[Q_1,Q_2]&=&-\int_0^t ds_1\int_0^{s_1} ds_2
[Q_1(s_1) -Q_2(s_1) ]\mu(s_1,s_2) [Q_1(s_2) +Q_2(s_2) ]  ,\nonumber \\ \relax
\Gamma_I[Q_1,Q_2]&=&\int_0^t ds_1\int_0^{s_1} ds_2  [Q_1(s_1) -Q_2(s_1)]
\nu(s_1,s_2) [Q_1(s_2) - Q_2(s_2) ] ,\nonumber \\ \relax
\label{tem}
\end{eqnarray}
where
 \begin{eqnarray}
\mu(\tau_1,\tau_2)&=&-\int_0^\infty\ d\omega \  {{\varepsilon^2}\over {2\pi
\omega}}
\sin{\omega(\tau_1-\tau_2)},
\nonumber \\ \relax
\nu(\tau_1,\tau_2)&=&\int_0^\infty\ d\omega \  {{\varepsilon^2}\over {2\pi
\omega}}
\coth{ {{\pi\omega}
\over a} } \cos{\omega(\tau_1-\tau_2)}
\label{kernel}
\end{eqnarray}
are respectively the noise and dissipation kernels arising for the same
detector at rest
in a  bath of thermal oscillators at temperature $ T=a/2\pi$ \cite{Fey1,CL}.

\section{Fluctuation in  acceleration}
\label{fluc}
In displaying the Unruh effect in the last section, it was assumed
that   the position degree of freedom of the detector was not quantum
mechanical, as the detector followed a determined
trajectory.    From a path integral point of view, this
treatment would require that the uniformly accelerating trajectory
sufficiently decohere from other neighboring paths to form a
consistent history.
 But of course the decoherence is always finite at a given time.
We now consider the uncertainty associated with    the fluctuations
in the detector's paths  around the uniformly accelerating trajectory.
For simplicity, we assume that
the value of the monopole  $Q$ is given by its quantum mechanical
average, supposed to be constant in time.
To study the extent to which the acceleration is well defined, we
consider  two uniformly accelerating trajectories with slightly different
accelerations:  $x_i=\sqrt{a_i^{-2}+t^2}$ with  $i=1,2$.
For the scalar field in  (\ref{mode}) with  interaction in (\ref{inte}) ,
the sum over modes can be converted into an integral, and  we then find
\cite{Demers}
\begin{equation}
\Gamma_I(a_1,a_2,t) =\frac{\varepsilon^2 Q^2}{8\pi}\int_0^t ds_1\int_0^{s_1}
ds_2\
P(s_1,s_2,a_1,a_2).
\label{def}
\end{equation}
where
\begin{equation}
 P(s_1,s_2,a_1,a_2)\equiv \ln \left(
 \frac{C^2-F^2}{A^2-F^2 }\ \ \frac{ D^2-F^2}{B^2-F^2 }
 \right)^2.
\label{P}
\end{equation}
and
\begin{eqnarray}
A&\equiv&\sqrt{a_1^{-2}+s_1^2}-\sqrt{a_1^{-2}+s_2^2},\hspace{2cm}
B\equiv\sqrt{a_2^{-2}+s_1^2}-\sqrt{a_2^{-2}+s_2^2},\nonumber \\ \relax
C&\equiv&\sqrt{a_1^{-2}+s_1^2}-\sqrt{a_2^{-2}+s_2^2}, \hspace{2cm}
D\equiv\sqrt{a_2^{-2}+s_1^2}-\sqrt{a_1^{-2}+s_2^2}, \nonumber \\ \relax
F&\equiv&s_1-s_2\ .
\end{eqnarray}
We now wish to extract the  features of $\Gamma_I$ for paths
lasting  a time large compared with the inverse accelerations
($t a_1\gg 1,
t a_2\gg 1$). Consider the rate of increase of $\Gamma_I$ with time.
{}From (\ref{def}),
\begin{equation}
{\dot{\Gamma}}_I (t,a_1,a_2)=\frac{\varepsilon^2 Q^2}{8\pi} t \ R(t,a_m,a_d),
\label{bu}
\end{equation}
with:
\begin{equation}
R(t,a_m,a_d)=
\int_0^1dy \
P(s_1=t,s_2=yt,a_1,a_2)
= R(\tilde{t}, \tilde{a}_d) .
\label{bi}
\end{equation}
In (\ref{bi}), we made the change of variable $s_2=yt$ ($y$
is unitless), and worked with the average $a_m\equiv
\frac{1}{2}(a_1^{-1}+a_2^{-1})$ and difference $a_d\equiv
(a_1^{-1}-a_2^{-1})$ in {\it inverse}  accelerations.
In the last line of (\ref{bi}), $a_m$ was used for the
scale factor in $P$, namely:  $\tilde{t}\equiv t/a_m$ and
$\tilde{a_d}\equiv a_d/a_m$.
Now except for a transient time $\tilde{t}{\ \lower-1.2pt\vbox{\hbox{\rlap{$<$}
\lower5pt\vbox{\hbox{$\sim$}}}}\ }1$,
we have
$ R(\tilde{t}, \tilde{a}_d)\ \rightarrow \ f(\tilde{a}_d) $
that is, $R$ is time independent  to a good approximation.
(Obviously, $f(\tilde{a}_d)$ is an even function.)
This can be seen in Fig.1, where numerical evaluations of $R$
are given  as a function of $\tilde{t}$ for a few representative values
of $\tilde{a}_d$.  (The apparently large value for small $\tilde{t}$ is damped
to a finite  and
narrow peak by the factor $t$ in (\ref{bu}).)
The function $f(\tilde{a}_d)$ may be evaluated numerically, and for small
$\tilde{a}_d$ is found to be nearly quadratic.
\begin{figure}
\hbox {\hspace{1cm} \resizebox{10cm}{!}{ \includegraphics*[40mm,140mm]
[175mm,230mm]{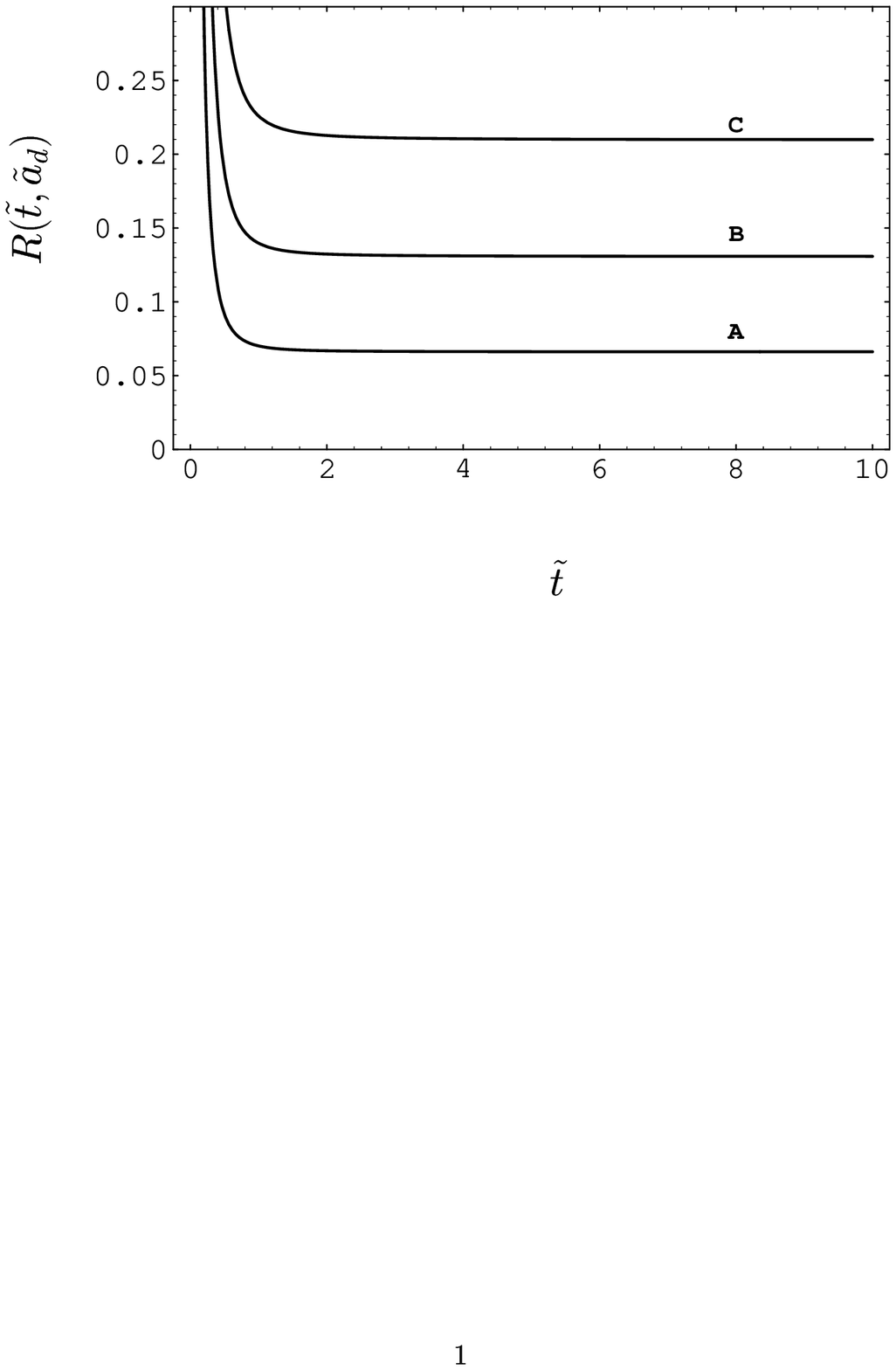}} }
\hbox{\hspace{1.5cm} \parbox{10cm}{\caption{
Plot of $R(\tilde{t},\tilde{a}_d)$
for $\tilde{a}_d=0.05$ (curve {\bf A}), $\tilde{a}_d=0.075$ (curve {\bf B})
and $\tilde{a}_d=0.1$ (curve {\bf C}). }}}
\end{figure}

With (\ref{bu}),
we thus have for $|\tilde{a}_d |\ll 1$ and $\tilde{t} \gg 1$ the excellent
approximation
\begin{equation}
{\Gamma}_I(t,a_1,a_2)\approx\frac{N}{16\pi} \varepsilon^2 \langle Q\rangle^2
 t^2 \left ( \frac{a_1-a_2}{(a_1+a_2)/2}\right ) ^2,
\label{deco}
\end{equation}
where $Q$ was replaced by its quantum mechanical average (quantum
fluctuations of $Q$ are neglected) and $N$ is a numerical constant.
We now take the condition $\Gamma_I \approx 1$ as  separating the decohering
and non-decohering accelerations.
Using (\ref{deco}) and the relation $ T=a/2\pi$ between temperature and
acceleration,
we conclude that a detector on a
decohered accelerating trajectory for a time $t$ will  be subject to
thermal fluctuations of the order:
\begin{equation}
\left | \frac{\Delta T}{T} \right |\approx \frac{1}{|\varepsilon | \langle
Q\rangle t}
\label{delta}
\end{equation}
where we dropped a constant of order one.

\section{Discussion} \label{Discussion}

Eq.  (\ref{delta})  can be  understood as an uncertainty relation between the
`time of
flight' of the detector  and the spread in temperature, with the
uncertainty given by $\frac{1}{|\varepsilon | \langle Q\rangle}$ and valid for
$t/T > 1$.
In Eq. (\ref{deco}), the quadratic dependence of the decoherence on
the coupling to the bath  is
general for a system coupled linearly to the bath
(by construction, the
amplitude of the  detector's excitations act to modify  the
effective coupling).
It is interesting to see that for the class of paths considered here,
the dependence  in time is also quadratic.
This shows that   the time is also making the coupling constant stronger
or inversely, that a larger coupling constant is identical to a longer
time evolution.
Qualitatively, this is in line with the result of Ref. \cite{CM}. As the
coupling to the bath gets stronger and that one waits longer, clearly
more particles are created and decoherence is
increased.
This can also be seen through (\ref{if2}). When particles are copiously
produced, a slight difference in the two paths will easily make  the
two states orthogonal.

Issues of decoherence similar to the ones considered here  could well
prove to be relevant in the
context of black holes. As it is well known,  one of the  distinguishing
features   of black holes is their  ability to Hawking radiate, in close
connection with the uniformly accelerating detector.  The environment
given  by these radiated particles could  then  serve in the  study
of the validity of the semi-classical approximation. This issue is currently
under investigation\cite{DK}.
\vskip 0.05in

\noindent
{\bf Acknowledgments.}
I   would like to thank   the  Organizing Committee for putting together
this interesting conference.
This work was supported in part by funds provided by the U.S. Department
of Energy under cooperative agreement \# DE-FC02-94ER40818.

\end{document}